\documentstyle[12pt]{article}
\linethickness{0.4pt}
\begin{document}
\thispagestyle{empty}
\title{Optimized perturbation methods\\for the free energy
of the anharmonic oscillator}
\author{Kostas Vlachos\\
Department of Physics, Patras University\\
GR 26110 Patras, Greece\\
\and
Anna Okopi\'nska\\
Institute of Physics, Warsaw University, Bia\l ystok Branch,\\
Lipowa 41, 15-424 Bia\l ystok, Poland\\e-mail: rozynek@fuw.edu.pl\\}
\maketitle
\begin{abstract}
\noindent Two possibile applications of the optimized expansion
for the free energy of the quantum-mechanical anharmonic
oscillator are discussed. The first method is for the finite
temperature effective potential; the second one, for the
classical effective potential. The results of both methods show
a quick convergence and agree well with the exact free energy
in the whole range of temperatures.
\end{abstract}
\clearpage
\noindent 1. The quantum mechanical anharmonic oscillator (AO)
with a Hamiltonian
\begin{equation}
H=\frac{p^2}{2} +\frac{m^2 x^2}{2} +\lambda x^4
\end{equation}
is equivalent~\cite{BW} to the theory of a scalar field in the
Euclidean space-time of one-dimension (time) with a classical
action given by
\begin{equation}
S[x]=\int\!\left[\frac{1}{2}x(t)(-\partial^2+m^2)x(t)+
\lambda x^{4}(t)\right]\,dt.
\label{Scl}
\end{equation}
After rescaling all quantities in terms of $\lambda$, only one
dimensionless parameter $z=\frac{m^2}{\lambda^{2/3}}$ remains;
therefore, when discussing numerical results, we put $\lambda=1$
without a loss of generality. The energy spectrum of the AO can
be calculated numerically and provides a simplest test for
approximation methods in quantum field theory (QFT) (f.e. loop
expansion~\cite{Jac}, 1/N expansion, loop expansion for
composite operators~\cite{CJT}, optimized expansion
(OE)~\cite{AO}). The conventional loop expansion gives the
Rayleigh-Schr\"odinger series which is asymptotic for the
AO~\cite{BW}; hence, the numerical results for energy levels are
good only for large values of the parameter
$\frac{m^2}{\lambda^{2/3}}$. In the $1/N$ expansion the energy
levels can be calculated from spectral properties of Green's
functions, derived from the given order effective action. The
results show a quick convergence and agree well with the exact
spectrum of the AO, the quality of approximation decreases for
increasing level of excitation~\cite{AOgf}. This is similar in
the OE; however, in this case the approximation becomes worse as
the parameter $\frac{m^2}{\lambda^{2/3}}$ decreases, becoming
unapplicable in the double-well case $(m^2<0)$~\cite{AOgf}.

It is also interesting to discuss the partition function and the
free energy which contain information on the whole spectrum of
the quantum system. The field-theoretical methods for systematic
approximations of the effective action can be extended to finite
temperature $T$ in the Euclidean formalism by compactifying the
"imaginary time" dimension with a period $\beta=\frac{1}{T}$.
The conventional loop expansion coincides with termodynamic
perturbation theory for free energy~\cite{Sch}. Here we study
the approximations for the free energy, obtained by two
applications of the OE: (i) for the finite temperature effective
potential, (ii) for the classical effective potential.\\

\noindent 2. The first method is an extension of the OE for the
effective action in QFT~\cite{AO} to finite temperature
effective potential and has been studied in the space-time of
arbitrary dimension~\cite{AOT}. Here we discuss the
one-dimensional case when the vacuum persistence amplitude is
defined by the integral
\begin{equation}
Z[J]=\int\!Dx\,e^{-S[x]+\int\!J(t)x(t)dt},
\label{Z}
\end{equation}
over the functions $x(t)$ which vanish at infinity. The
effective action is defined by
\begin{equation}
\Gamma[\xi]=\ln Z[J]-\int\!\xi(t)J(t)dt,
\label{Gam}
\end{equation}
where the expectation value of $x$ is given by
$\xi(t)=\frac{\delta \ln Z}{\delta J(t)}$. The extension to
finite temperature $T$ consists in replacing the set of functions
in the integral~(\ref{Z}) by periodic functions with a period
$\beta=\frac{1}{T}$. The infinite interval in all integrals over
$t$ has to be replaced by the interval $[0,\beta]$, we denote
the corresponding quantities by $ Z_{\beta}, \Gamma_{\beta}$.
The finite temperature effective potential defined by
\begin{equation}
V_{\beta}(\xi)=\frac{1}{\beta}\left.\Gamma_{\beta}[\xi]\right|_{\xi(t)=\xi}
\label{FTEP}
\end{equation}
has a meaning of the free energy of the quantum system
interacting with a constant electric field $J$, when the
expectation value of the coordinate $x$ is given by
$\xi=const$. The free energy $F$ of the AO is determined by the
value of the finite temperature effective potential at minimum
which corresponds to $J=0$.

The OE for the effective action is generated by the application
of the steepest-descent method to the generating functional
$Z[J]$ with the classical action modified to the form
\begin{eqnarray}
S_{\epsilon}[x]&=&\int\frac{1}{2}x(t)G^{-1}(t,t')x(t')dt dt'\nonumber\\
&+&\epsilon\left[\int\frac{1}{2}x(t)(D^{-1}(t,t')-G^{-1}(t,t'))x(t')dt
dt'+\lambda\int x^{4}(t)dt\right],
\label{Seps}
\end{eqnarray}
where $D^{-1}(t,t')=(-\partial^2+m^2)\delta (t-t')$. After
shifting $x(t)$ by $x_{0}(t)$ chosen to satisfy the classical
equation of motion $\frac{\delta S_{\epsilon}}{\delta
x_{0}(t)}=-J(t)$ and expanding the exponential into Taylor
series we obtain
\newpage
\begin{eqnarray}
Z[J]\!&=&\!e^{-S_{\epsilon}[x_{0}]+\int\!J(t)x_{0}(t)dt}
\int\!Dx'e^{-\int
\frac{1}{2}x'(t)G^{-1}(t,t')x'(t)dt dt'}\nonumber\\
\times [1&-&\epsilon\,(\int\!\frac{1}{2}\!
x'(t)(\Delta^{-1}(t,t')-G^{-1}(t,t'))x'(t')dt dt'\!+\lambda
\int x_{0}(t)x'^3(t) dt\nonumber\\&+&\lambda \int x'^{4}(t) dt\,)+...\,],
\label{Zeps}
\end{eqnarray}
where $\Delta^{-1}(t,t')=(-\partial^2+m^2+12\lambda
x_{0}^2(t))\delta (t-t')$. Upon Legendre transform, the effective
action~(\ref{Gam}) is obtained as a series in $\epsilon$. The
$n$th order term can be represented diagramatically as a sum of
$n$-vertex vacuum one-particle-irreducible diagrams with Feynman
rules of the modified theory~(\ref{Seps}). The third order
result is shown in Fig.1.\\

\noindent 3. The finite temperature effective potential in the OE is given
by the same set of diagrams as the effective action, only the
Feynman rules are replaced by those at finite temperature. Since
$V_{\beta}$ is a function of constant $\xi$, the propagator can
be chosen in the form
\begin{equation}
G_{\beta}(t,t')=\frac{1}{\beta}\Sigma_{m=-\infty}^{\infty}
\exp[-i\omega_{m}(t-t')] \frac{1}{\omega_{m}^{2}+\Omega^2}=
\frac{\cosh[\frac{\Omega}{2}(|t-t'|-\beta)]}
{\Omega \sinh[\frac{\beta\Omega}{2}]},
\label{Gbeta}
\end{equation}
with an arbitrary parameter $\Omega$. The Matsubara frequencies
are given by $\omega_{m}=\frac{2\pi m}{\beta}$. At zero
temperature the propagator becomes equal to
\begin{equation}
G_{\infty}(t,t')=\int \frac{dp}{2\pi}\exp[-ip(t-t')]
\frac{1}{p^2+\Omega^2}=
\frac{\exp[-\Omega(|t-t'|]}{2\Omega}.
\end{equation}

The parameter $\epsilon$ is a formal parameter of expansion and
is set equal to one at the end. The exact $V_{\beta}(\xi)$,
obtained as a sum of an infinite series, does not depend on
$\Omega$, but a finite order truncation does. We can make the
$n$th-order approximant $V_{\beta}^{(n)}(\Omega,\xi)$ as
insensitive as possible to small variation of the unphysical
parameter by choosing $\Omega$ equal to $\tilde{\Omega}(n,\xi,\beta)$
satisfying the gap equation
\begin{equation}
\frac{\delta V_{\beta}^{(n)}}{\delta\tilde{\Omega}}= 0.
\label{sta}
\end{equation}
In the approximate expression for the finite temperature
effective potential
$V_{\beta}^{(n)}(\xi)=V_{\beta}^{(n)}(\xi,\tilde{\Omega}(n,\xi,\beta))$
the optimal value of $\Omega$ changes from order to order,
assuring the convergence of the expansion~\cite{conv}. The first
order of the OE coincides with the finite temperature Hartree
approximation, but the variational interpretation cannot by
maintained beyond the first order.

We calculated the finite temperature effective potential to
third order in the OE. In second order the gap equation has no
real solution in some range of temperatures, in this case the
real part of the result has been taken. For the single-well AO
$(m^2>0)$ the symmetric minimum of the finite temperature effective
potential at $\xi=0$ (OES) gives a very good description of the free
energy. The quality of approximation becomes worse for
decreasing $\frac{m^2}{\lambda^{2/3}}$. In the most unfavourable
case of the quartic oscillator $(m^2=0)$ the results of three
lowest orders are shown in Fig.~2 in comparison with the exact
free energy, calculated by the numerical procedure based on the
modification of the linear variational method~\cite{AOAO}.

In the double-well case $(m^2<0)$, there is a critical value
$z_{cr}(n,\beta)$ of the parameter $\frac{m^2}{\lambda^{2/3}}$,
above which the only minimum of $V_{\beta}^{(n)}(\xi)$ is at
$\xi=0$, in agreement with the exact result. However, for
$\frac{m^2}{\lambda^{2/3}}$ below $z_{cr}$ a lower minimum at
$\xi\ne 0$ appears. As can be seen in Fig.~3, for $m^2=-20$, the
value of the finite temperature effective potential in the
non-symmetric minimum (OEN) gives much better approximation of
the free energy than the value at $\xi=0$ (OES).\\

\noindent 4. Different approximation schemes for the free energy can be
obtained~\cite{Klei} approximating the classical effective potential
$V_{cl}(x)$, defined by a simple integral
\begin{equation}
Z_{\beta}[J=0]=e^{-\beta F}=\int\! \frac{dx_{0}}{\sqrt
(2\pi\beta)}e^{-\beta V_{cl}(x_{0})}.
\label{Vcl}
\end{equation}
The local partition function $Z^{x_{0}}$ can be written as
\begin{equation}
Z^{x_{0}}=e^{-\beta V_{cl}(x_{0})}= \int \! Dx\,\sqrt
(2\pi\beta )\delta(x_{0}-\overline{x}) e^{- S[x]},
\label{Zloc}
\end{equation}
where $\overline{x}=\frac{\int\!d\tau x(\tau)}{\beta}$. For
$T=0$ the classical effective potential coincides with the
constraint effective potential~\cite{CEP}.

The OE for the classical effective potential is generated by the
application of the steepest-descent method to
$Z^{x_{0}}$~(\ref{Zloc}) with the modified classical
action~(\ref{Seps}). After shifting $x(t)$ by a constant $x_{0}$
and expanding the exponential into the Taylor series we have
\begin{eqnarray}
Z^{x_{0}}&=&e^{-\beta\left(\frac{\Omega^2}{2}x_{0}^2+
\epsilon\left(\frac{1}{2}(m^2-\Omega^2)
x_{0}^2+\lambda x_{0}^{4}\right)\right)}\nonumber\\
&\times&\int\!Dx'\!\sqrt (2\pi\beta
)\delta(x_{0}-\overline{x})e^{-\int_{0}^{\beta}
\frac{1}{2}x'(t)(-\partial^2+\Omega^2)x'(t)\,dt}\nonumber\\
&\times&[1-\epsilon\,(\frac{m^2\!+\!12\lambda
x_{0}^2-\Omega^2}{2}\int_{0}^{\beta} x'(t)^2\,dt
+\lambda x_{0}\int_{0}^{\beta} x'(t)^3\,dt\nonumber\\
&+&\lambda \int_{0}^{\beta} x'^{4}(t)\,dt\,)+...\,].
\label{Zleps}
\end{eqnarray}
After performing the Gaussian integrals, the local partition
function and the classical effective potential can be calculated
as series in $\epsilon$. The given order classical effective
potential $V_{cl}^{n}(x_{0})$ in the OE can be represented
diagramatically, only one-particle irreducible diagrams are
present, because by definition $x'(t)$ has zero vacuum
expectation value. Therefore, the set of diagrams is the same as
in the case of the OE for the finite temperature effective
potential, the only difference is in the propagator. In the case
of the OE for the classical effective potential the propagator
does not contain zero modes and equals
\begin{equation}
G_{\beta}^{wzm}(t,t')=\frac{2}{\beta}\Sigma_{m=1}^{\infty}
\exp[-i\omega_{m}(t-t')] \frac{1}{\omega_{m}^{2}+\Omega^2}
=G_{\beta}(t,t')-\frac{1}{\beta\Omega^2}
\end{equation}
where $G_{\beta}(t,t')$ is the finite temperature
propagator~(\ref{Gbeta}). Performing the integration over
$x_{0}$ in~(\ref{Vcl}) and calculating the free energy to the
given order in $\epsilon$ we would obtain the result coincident
with the OE for the finite temperature effective potential.
However, a better approximation for the partition function can
be obtained by reversing the order of operation: first to
optimize the given order classical effective potential
\begin{equation}
\frac{\delta V_{cl}^{(n)}}{\delta\Omega} = 0,
\label{staFK}
\end{equation}
and after to perform the integration over $x_{0}$ in~(\ref{Vcl})
numerically. The first order of this approximation coincides
with the Feynman-Kleinert (FK) variational determination of the
classical effective potential~\cite{GT,FK}. The OE gives a
possibility to calculate corrections to the variational
classical effective potential, improving the FK approximation
for partition function in a systematic way. For alternative
methods of calculating corrections to FK approximation see
Refs.~\cite{Klei,JK}.\\

\noindent 5. We have studied the numerical results of the OE for the
classical effective potential in the first and third order (to
avoid complications with complex solutions of the second
order). For $m^2\le 0$ the quality of the approximation becomes
worse for decreasing $\frac{m^2}{\lambda^{2/3}}$; the results
for the quartic oscillator $(m^2=0)$ are compared with the exact
free energy in Fig.~4. The first order results (FK1) are better
than obtained from the value of the finite temperature effective
potential at $\xi=0$ (OES1); however, the third order results
(FK3) and (OES3) are very similar and agree well with the exact
free energy calculated numerically. The differencies between the
studied methods decrease to zero for $T=0$.

In the double-well case, the results of the OE for the classical
effective potential are better than obtained from the lowest
minimum (at $\xi\ne 0$) of the effective potential (OEN). The
quality of the approximation improves for increasing
$\frac{|m^2|}{\lambda^{2/3}}$. For $m^2=-20$ the first order
results (FK1) are indistinguishable from the exact results on
the scale of Fig.~3. In Fig.~5 we compare the results of both
methods for larger range of temperatures, the convergence in the
FK approach is better than in the case of the OEN. However, one
has to notice that the numerical calculations are much more
complicated in the OE for the classical effective potential.
Moreover, a generalisation of the FK approach to the true QFT
(with an untrivial space dimension) is difficult, even in the
lowest order.\\

\noindent {\bf{\Large Acknowledgements}}\\

This work has been supported partially by the Comitee for
Scientific Research under Grant PB-2-0956-91-01. One of the
autors (K.V.) is grateful to the members of the Institute of
Physics (Bia\l ystok Branch of Warsaw University) and So\l tan
Institute for Nuclear Studies for the kind hospitality.

\newpage
\noindent {\bf{\Large Figure captions}}\\

\noindent Figure 1. The effective action~to third order of the
optimized expansion. The full line is an arbitrary propagator
$G(t,t')$, the rule denotes the two-particle vertex
$\Delta^{-1}(t,t')-G^{-1}(t,t')$.\\

\noindent Figure 2. The free energy $F$ of the quartic oscillator
$(m^2=0)$, obtained as the value of the finite temperature
effective potential~at $\xi=0$ in first three orders (OES1,
OES2, OES3) of the OE, plotted {\em vs.} the inverse temperature
$1/T$.\\

\noindent Figure 3. The free energy $F$ of the double-well
oscillator $(m^2=-20)$, obtained in the OE as the value of the
effective potential~at $\xi=0$ (OES), and at $\xi\ne 0$ (OEN).\\

\noindent Figure 4. The free energy $F$ of the quartic oscillator
$(m^2=0)$, calculated with the classical effective potential
in first (FK1) and third (FK3) orders of the OE.\\

\noindent Figure 5. The free energy $F$ of the double-well
oscillator $(m^2=-20)$, calculated with the classical effective
potential in first (FK1) and third (FK3) orders of the OE,
compared with the value of the finite temperature effective
potential at $\xi\ne 0$ (OEN1, OEN3).
\end{document}